\documentclass[traditabstract]{aa}
\usepackage{natbib}
\usepackage{txfonts}
\usepackage{graphicx}
\titlerunning{Limb darkening in FGK dwarf stars}
\authorrunning{H.~R. Neilson}

\begin{document}

\title{Spherically symmetric model stellar atmospheres and limb darkening II: 
limb-darkening laws, gravity-darkening coefficients and angular 
diameter corrections for FGK dwarf stars\thanks{Tables 2 --17 are only 
available in electronic form at the CDS via anonymous ftp to 
cdsarc-u-strasbg.fr or via http://cdsweb.u-strasbg.fr/cgi-bin/qcat?J/A+A/}}

\author{Hilding R. Neilson\inst{1} \and John B. Lester \inst{2,3}}
\institute{
   Department of Physics \& Astronomy, East Tennessee State University, 
   Box 70652, Johnson City, TN 37614 USA
   \email{neilsonh@etsu.edu}
 \and 
     Department of Chemical and Physical Sciences, 
     University of Toronto Mississauga 
 \and
     Department of Astronomy \& Astrophysics, University of Toronto \\
     \email{lester@astro.utoronto.ca}
  }

\date{}

\abstract{
Limb darkening is a fundamental ingredient for interpreting observations of 
planetary transits, eclipsing binaries, optical/infrared interferometry and microlensing events.  
However, this modeling traditionally represents limb darkening by a simple 
law having one or two coefficients that have been derived from plane-parallel 
model stellar atmospheres, which has been done by many researchers.  
More recently, researchers have gone beyond plane-parallel models and 
considered other geometries.
We 
previously studied the limb-darkening coefficients from spherically symmetric 
and plane-parallel model stellar atmospheres for cool giant and supergiant 
stars, and in this investigation we apply the same techniques to FGK dwarf 
stars.  We present limb-darkening coefficients, gravity-darkening coefficients 
and interferometric angular diameter corrections from \textit{Atlas} and 
\textit{SAtlas} model stellar atmospheres.  
We find that sphericity is important even for dwarf model atmospheres, leading 
to significant differences in the predicted coefficients.
}
\maketitle

\section{Introduction}
One of the great astronomical advances of the past two decades has been the discovery and study of 
extrasolar planets via the transit method, i.e.~from the minute drop of a 
star's flux as a planet passes in front of it.  The transit not 
only constrains the planet's properties but also the star's properties, such as limb
darkening.  However, interpreting planetary transits typically assumes 
that limb darkening can be parametrized by a simple relation 
\citep{Mandel2003} with a few free parameters that can be fit directly from the
observations or assumed from model stellar atmospheres. 

Limb darkening is important not only for understanding planetary transits 
\citep[e.g.][]{Croll2011}, but also for interpreting optical interferometric 
observations \citep[e.g.][]{Davis2000} and microlensing observations 
\citep[e.g.][]{An2002} and eclipsing binary light curves \citep[e.g.][]{Bass2012}.  Like transit measurements, interferometric and 
microlensing observations are typically fit by simple limb-darkening laws 
with coefficients derived from model stellar atmospheres 
\citep{Al-naimiy1979, Wade1985, vanhamme1993,Claret2000,Claret2011, 
Claret2012}.  However, these simple limb-darkening laws have become less 
suitable as the observations have improved.  For example, \cite{Fields2003} showed that 
flux-normalized limb-darkening laws fit to {\it Atlas} plane-parallel model 
atmospheres disagreed with microlensing observations.  Limb-darkening 
coefficients derived from planetary transit observations with large impact 
parameters differ more from the limb-darkening coefficients from model 
atmospheres, but the discrepancy still exists when the impact parameter is 
taken into account \citep{Barros2012}.

This discrepancy might be due to a number of physical processes, including 
granulation, multidimensional convection and/or the presence of magnetic 
fields in the stellar atmosphere.  However, the simplest step is to 
assume a more realistic geometry for the model stellar atmospheres.  
Limb-darkening coefficients presented in the literature are based on two 
forms: plane-parallel model stellar atmospheres computed using the {\it Atlas} 
\citep{Kurucz1979} and {\it Phoenix} code \citep{Hauschildt1998} and 
spherically symmetric model stellar atmospheres also computed from the 
{\it Phoenix} code \citep{Sing2010, Howarth2011a, Claret2011, Claret2012}.  In particular, \citet{Claret2003}  and \citet{Claret2012, Claret2013} explored limb darkening 
using spherically symmetric {\it Phoenix} model stellar atmospheres specifically for main 
sequence stars. 
They also introduced the concept of ``quasi-spherical'' models, defined as the spherically-symmetric intensity profile restricted to inner part of the stellar disk ($\mu \geq 0.1$), to compare limb-darkening coefficients with those from plane-parallel models.  

In our previous study    \citep[][hereafter Paper~1]{Neilson2012}, we presented
               coefficients for six typical limb-darkening laws fit to the surface intensities 
for grids of plane-parallel and spherical model atmospheres \citep{Lester2008} 
representing red giant and supergiant stars.  The intensities were for the wavebands of the Johnson-Cousins 
\citep{Johnson1953, Bessell2005}, {\em CoRot} \citep{Auvergne2009} and 
{\em Kepler} \citep{Koch2004} filters.  We also computed gravity-darkening 
coefficients and interferometric angular diameter corrections. We found that the predicted 
limb-darkening coefficients computed from spherical model atmospheres differ 
from those computed from plane-parallel model atmospheres, which was 
not unexpected; the height of the atmospheres of red giants and supergiants 
relative to the stellar radius is many times greater than the relative heights of 
the atmospheres of main-sequence stars, i.e. the assumed model geometry is 
important.  We found similar differences for the angular diameter corrections 
as a function of geometry but little difference between gravity-darkening 
coefficients as a function of geometry.  While model atmosphere geometry is 
clearly important for understanding the extended atmospheres of red giant and 
supergiant stars, it is not as obvious that   geometry also changes predictions 
for model stellar atmospheres of main sequence dwarf stars \citep[e.g.][]{Claret2003}.

In this work, we explore the role of model atmosphere geometry in understanding
limb darkening in dwarf stars and compute limb-darkening coefficients, 
gravity-darkening coefficients and interferometric angular diameter 
corrections from grids of model stellar atmospheres of dwarf stars.  In 
Sect.~2, we briefly describe the grids of model atmospheres used.  In Sect.~3,
we describe various limb-darkening laws and compare predicted limb-darkening 
coefficients, while in Sect.~4 we compute gravity-darkening coefficients.  We 
present interferometric angular diameter corrections as a function of geometry 
in Sect.~5 and summarize our results in Sect.~6.

\section{Model stellar atmospheres}
\begin{figure}[t]
\begin{center}
\includegraphics[width=0.5\textwidth]{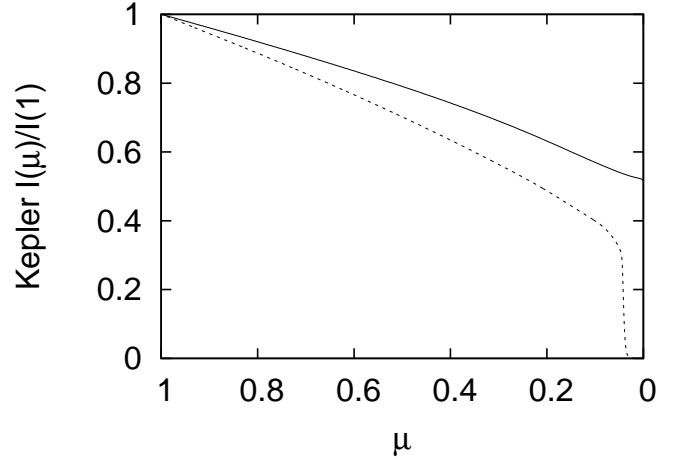}
\end{center}
\caption{
Predicted {\em Kepler}-band intensity profiles for plane-parallel 
(solid line) and spherically symmetric (dotted line) model stellar 
atmosphere with $T_{\rm{eff}} = 5800~$K, $\log g = 4.5$ and $M= 1.1~M_\odot$.}
\label{f0}
\end{figure}

The {\it Atlas/SAtlas} code was used to compute model stellar atmospheres 
assuming either plane-parallel or spherically symmetric geometry.  Details of 
the code can be found in \cite{Lester2008}, \cite{Neilson2011, Neilson2012a} 
and Paper~1.  We computed model stellar atmospheres with parameters 
$3500$~K $\leq T_{\rm{eff}} \leq 8000$~K in steps of $100~$K, and 
$4.0 \leq \log g \leq 4.75$ in steps of $0.25$.  For the spherically symmetric 
models, which require an additional parameter, such as stellar mass, to characterize the atmosphere, 
we chose $M = 0.2$ to $1.4~M_\odot$ in 
steps of $0.3~M_\odot$.  For each model stellar atmosphere we compute 
intensities at each wavelength for 1000 uniformly spaced values of $\mu$, the cosine of the angle formed by the 
line-of-sight point on the stellar disk and the disk center, 
spanning $0 \leq \mu \leq 1$. Typically, 
{\it Atlas} models compute intensities at only seventeen angles 
\citep{Kurucz1979}, but some have employed 100 $\mu$-points \citep{Claret2011}.
We compute intensity profiles for each model atmosphere for the $BVRIH$ and 
$K$-bands as well as the {\em CoRot} and {\em Kepler}-bands. As an example, 
Fig.~\ref{f0} shows the {\em Kepler}-band intensity profiles for 
plane-parallel and spherical models with 
$T_\mathrm{eff} = 5800~$K, $\log g = 4.5$ and $M = 1.1~M_\odot$.
Using the wavebands outlined above,         we compute limb-darkening 
coefficients, gravity-darkening coefficients and interferometric angular 
diameter corrections.

\section{Limb-darkening laws}
\begin{figure}[t]
\begin{center}
\includegraphics[width=0.5\textwidth]{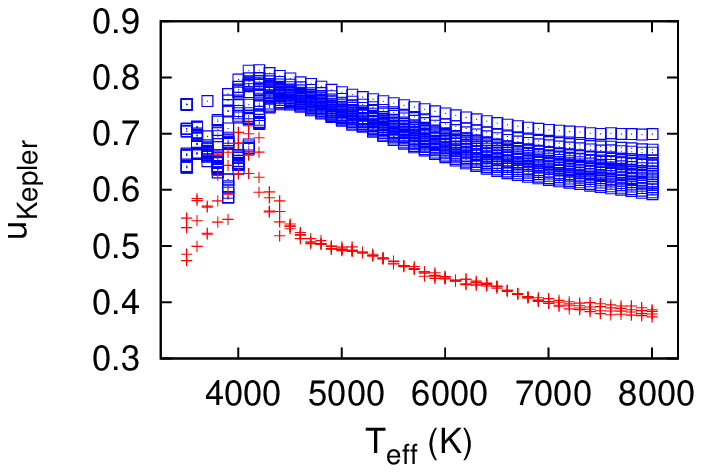}
\end{center}
\caption{The limb-darkening coefficient $u$, used in the linear law (Eq.~\ref{eq:linear}),
 applied to the \textit{Kepler} photometric band. Crosses are the 
plane-parallel model stellar atmospheres, and the squares are the 
spherical models.}
\label{fig:ab_l}
\end{figure}

\begin{figure*}[t]
\begin{center}
\includegraphics[width=0.5\textwidth]{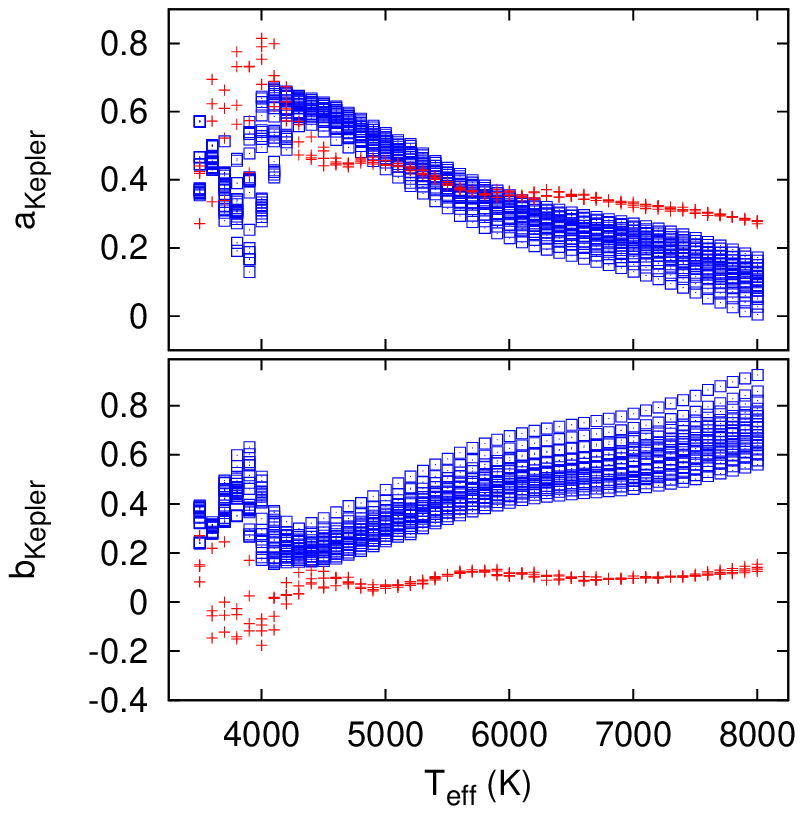}\includegraphics[width=0.5\textwidth]{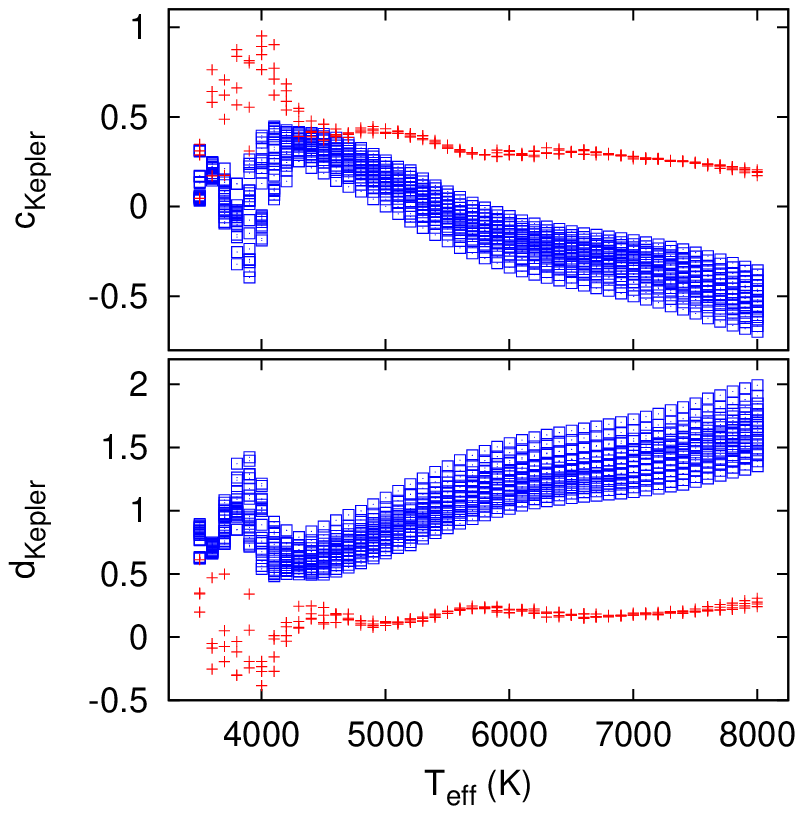}
\end{center}
\caption{Limb-darkening coefficients $a$ and $b$ used in the quadratic law 
(Eq.~\ref{eq:quad}) (left panel), and the coefficients $c$ and $d$ used in the 
square-root law (Eq.~\ref{eq:root}) (right panel), all applied to the \textit{Kepler} 
photometric band.  The symbols have the same meanings as in 
Fig.~\ref{fig:ab_l}. }\label{f2}
\end{figure*}

\begin{figure*}[t]
\begin{center}
\includegraphics[width=0.5\textwidth]{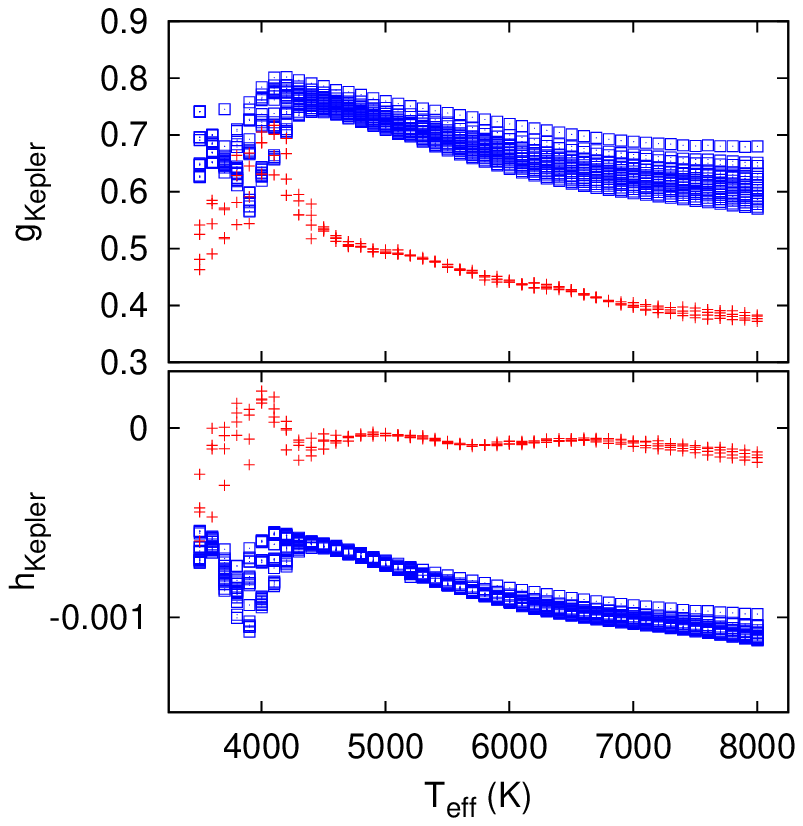}\includegraphics[width=0.5\textwidth]{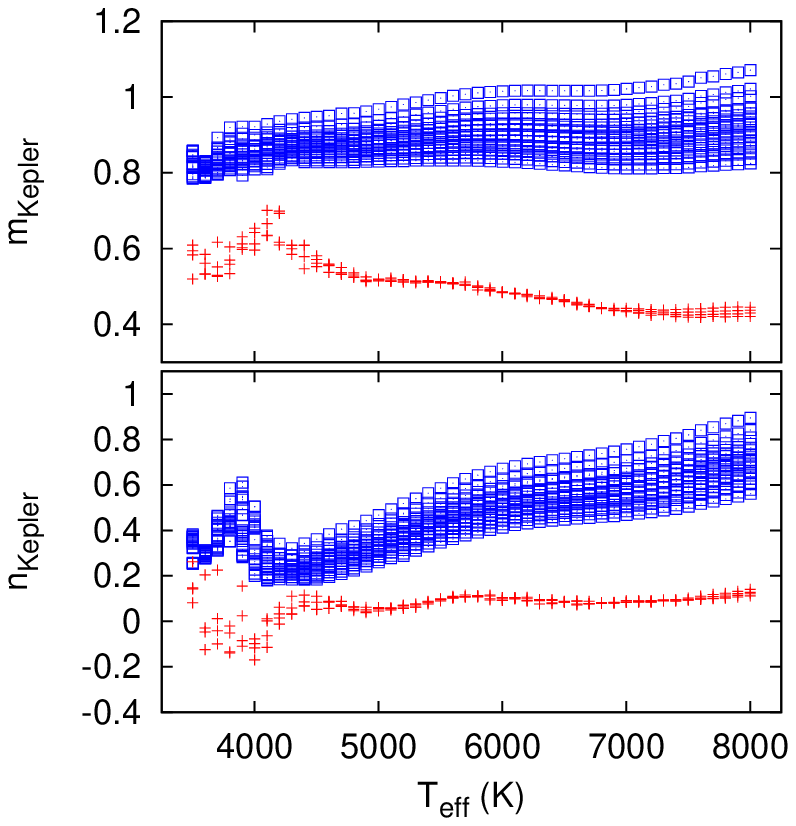}
\end{center}
\caption{Limb-darkening coefficients $g$ and $h$ used in the exponential law 
(Eq.~\ref{eq:exp}) (left panel), and the coefficients $m$ and $n$ used in the 
logarithmic law 
(Eq.~\ref{eq:ln}) (right panel), all applied to the \textit{Kepler} photometric 
 band.  The symbols have the same meanings as in Fig.~\ref{fig:ab_l}. }\label{f3}
\end{figure*}

\begin{figure*}[t]
\begin{center}
\includegraphics[width=0.5\textwidth]{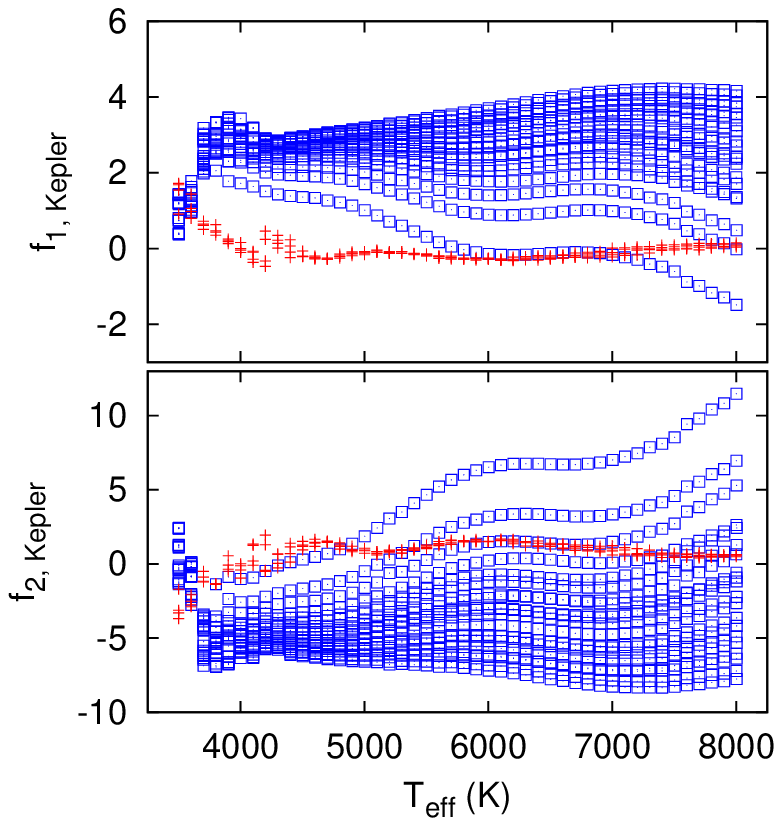}\includegraphics[width=0.5\textwidth]{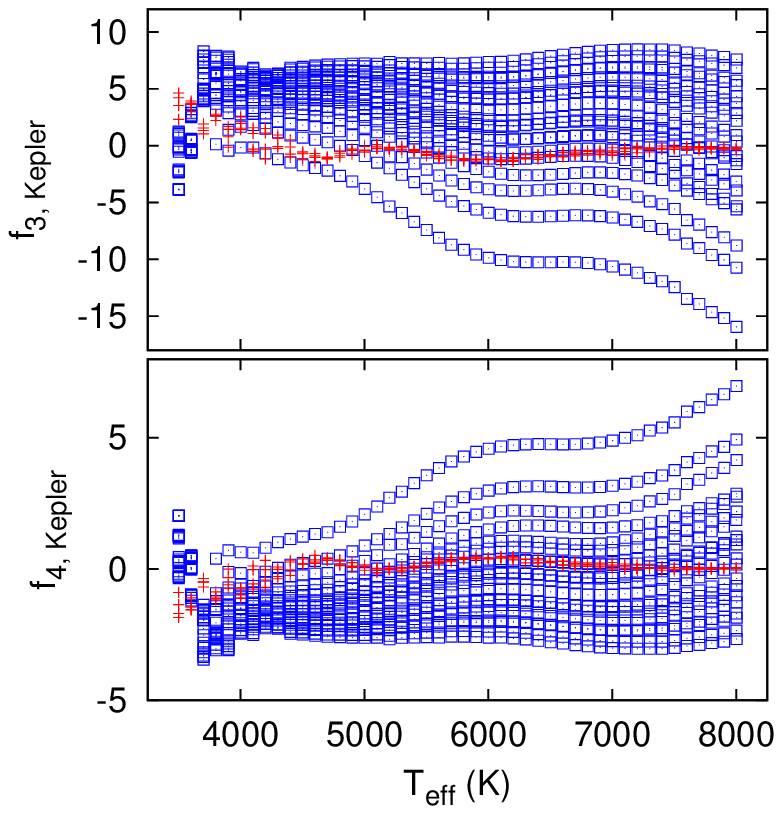}
\end{center}
\caption{Limb-darkening coefficients $f_1$, $f_2$, $f_3$ and $f_4$ used in the 
\citet{Claret2000} four-parameter law, Eq.~\ref{eq:4-p}, applied to the 
\textit{Kepler} photometric band.  The symbols are the same as in 
Fig.~\ref{fig:ab_l}. }\label{f4}
\end{figure*}

We consider the same six limb-darkening laws as in Paper~1:
\begin{equation}\label{eq:linear}
\frac{I(\mu)}{I(\mu=1)} = 1 - u(1- \mu) \hfill \makebox{Linear,\hspace{.3cm}}
\end{equation}
\begin{equation}\label{eq:quad}
\frac{I(\mu)}{I(\mu=1)} = 1 - a(1-\mu) - b(1-\mu)^2 \hfill\mbox{Quadratic,\hspace{.3cm}}
\end{equation}
\begin{equation}\label{eq:root}
\frac{I(\mu)}{I(\mu=1)} = 1 - c(1-\mu) - d(1-\sqrt{\mu}) \hfill \mbox{Square-Root,\hspace{.3cm}}
\end{equation}
\begin{equation}\label{eq:4-p}
\frac{I(\mu)}{I(\mu=1)} = 1 - \sum_{j=1}^{4}f_j(1-\mu^{j/2}) \hfill \mbox{4-Parameter,\hspace{.3cm}}
\end{equation}
\begin{equation}\label{eq:exp}
\frac{I(\mu)}{I(\mu=1)} = 1 - g(1-\mu) - h\frac{1}{1-e^{\mu}} \hfill \mbox{Exponential,\hspace{.3cm}}
\end{equation}
\begin{equation}\label{eq:ln}
\frac{I(\mu)}{I(\mu=1)} = 1 - m(1-\mu) - n\mu \ln \mu \hfill \mbox{Logarithmic.\hspace{.3cm}}
\end{equation}
As in Paper~1, we use a general least-squares fitting algorithm to compute the limb-darkening coefficients for each law in the 
$BVRIH$- and $K$-bands as well as for the {\em CoRot} and {\em Kepler}-bands. 
Using the \textit{Kepler}-band as an example, 
Figure~\ref{fig:ab_l} shows the best-fit limb-darkening coefficient 
for the linear law (Eq.~\ref{eq:linear}), Fig.~\ref{f2} shows the coefficients for the quadratic (Eq.~\ref{eq:quad}) and square-root (Eq.~\ref{eq:root}) laws,
Fig.~\ref{f3} shows the coefficients for the exponential (Eq.~\ref{eq:exp}) and logarithmic (Eq.~\ref{eq:ln}) laws and Fig.~\ref{f4} shows the coefficients 
for the \cite{Claret2000} four-parameter law (Eq.~\ref{eq:4-p}).


The results shown in Fig.~\ref{fig:ab_l} demonstrate how the geometry of the 
model atmosphere affects the best-fit linear {\em Kepler}-band limb-darkening 
coefficients, with squares representing fits to the spherically symmetric model
atmospheres and crosses representing fits to the plane-parallel models.  The values of the 
$u$-coefficient for the spherical models are larger than those for the 
     planar models, particularly for models with $T_{\rm{eff}} > 4500~$K.  At 
these higher effective temperatures the difference due to geometry, 
$\Delta u_{\rm{Kepler}}$, is $\sim 0.3$.  There is also a greater
spread for the spherical model coefficients at a given effective temperature.  This 
is caused by the spherical models being defined by three parameters, with mass and 
radius being separated, as opposed to the 
two parameters for plane-parallel model atmospheres, where mass and radius are 
combined in the surface gravity. 
               At $T_{\rm{eff}} < 4500~$K the 
$u$-coefficients computed for both geometries shift to similar values.  A 
likely cause of this change relative to the higher effective temperatures 
is the shift in dominant opacities from H$^-$ to TiO.

The more complex limb-darkening laws have similar differences between 
coefficients from plane-parallel and spherically symmetric models.  For the 
quadratic and square-root laws, the coefficients of the linear term ($a$ and 
$c$, respectively) shows similar behavior as a function of effective 
temperature as does the $u$-coefficients, while the coefficients of the 
non-linear terms ($b$ and $d$) appear correlated to the coefficients of the 
linear terms, as was seen previously for other laws \citep{Fields2003, 
Neilson2011, Neilson2012a}.

For the exponential and logarithmic laws, the best-fit coefficients again 
differ as a function of model atmosphere geometry. The limb-darkening 
coefficients also appear to be correlated for each law. It is notable that the 
best-fit $m$-coefficients of the logarithmic law from spherically symmetric 
models are approximately constant with respect to effective temperature, 
whereas the non-linear term is not constant. The limb-darkening coefficients 
from spherically symmetric models for both exponential and logarithmic laws 
vary significantly for any given effective temperature, suggesting the 
coefficients are sensitive to the mass and gravity of a model stellar 
atmosphere.

The best-fit coefficients for the \cite{Claret2000} four-parameter 
limb-darkening laws do not agree for spherical and plane-parallel models.  For 
effective temperatures greater than $4000$~K, the limb-darkening coefficient 
$f_1$ varies from $-2$ to $+4$ for the spherical models but only from $-0.5$ to $0.5$ 
for the plane-parallel models.  The dramatic difference is due to the more     
complex structure of spherically symmetric model intensity profiles, even when 
considering the smaller atmospheric extensions for models used in this work as 
opposed to those considered in Paper~1, which indicates that even this 
more sophisticated limb-darkening law is not ideal for fitting spherically 
symmetric model intensity profiles.

\begin{figure*}[t]
\begin{center}
\includegraphics[width=0.5\textwidth]{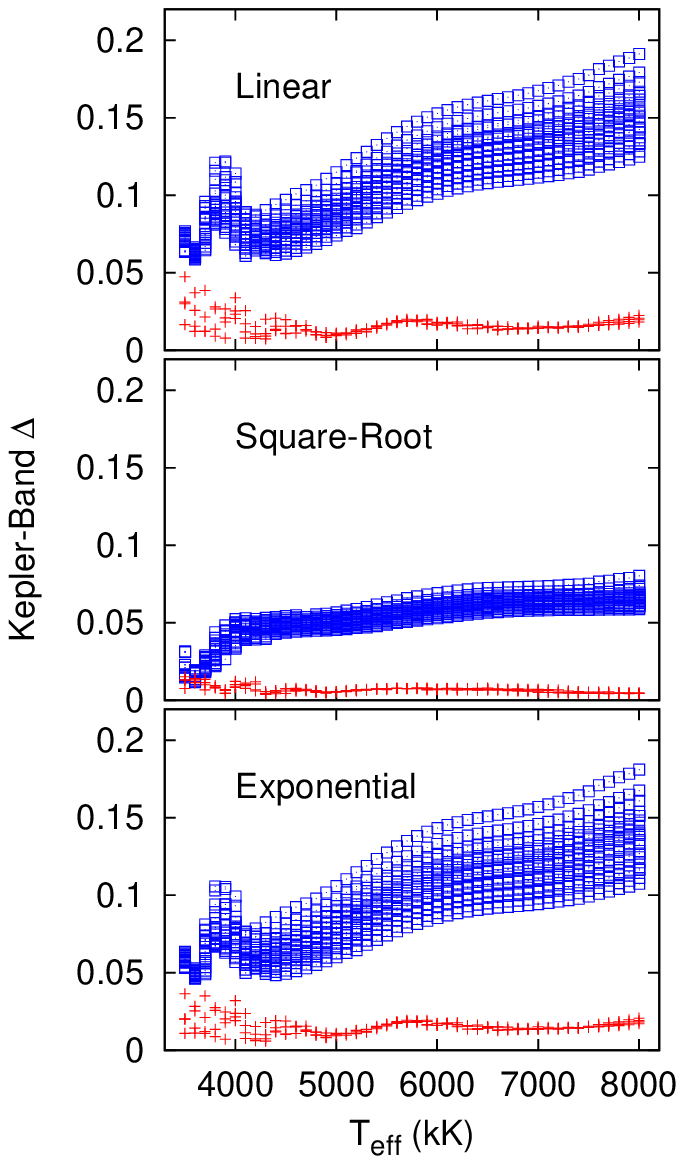}\includegraphics[width=0.5\textwidth]{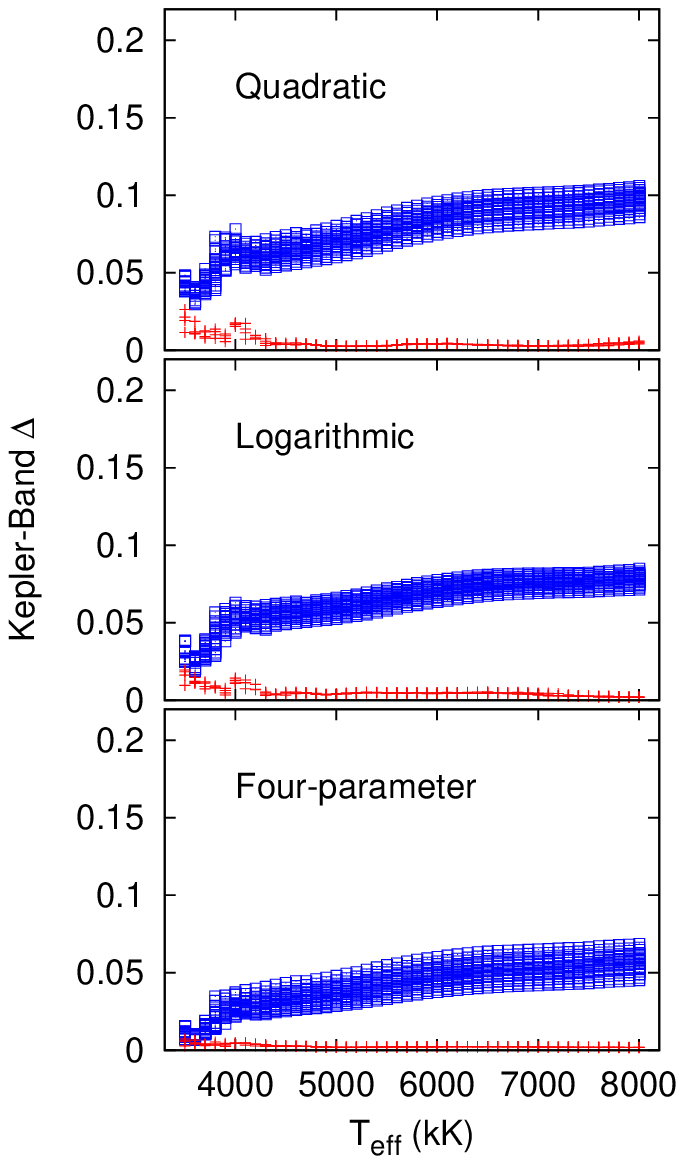}
\end{center}
\caption{The error of the best-fit limb-darkening relation, defined by 
Eq.~\ref{eq:delta}, for every model atmosphere (crosses represent 
plane-parallel models, squares spherical models) for each of the 
six limb-darkening laws at {\it Kepler}-band wavelengths.}\label{f5}
\end{figure*}

Figures~\ref{fig:ab_l}--\ref{f4} demonstrate that the best-fit coefficients from spherical models differ from those computed 
from planar models, but these figures do not quantify the fits for either geometry.
To do this, we employ the parameter defined in Paper~1, 
\begin{equation}\label{eq:delta}
\Delta_{\lambda} \equiv \sqrt{ \frac{\sum \left [I_\mathrm{model}(\mu) - I_\mathrm{fit}(\mu) \right ]^2}{\sum \left [I_\mathrm{fit}(\mu) \right ]^2}},
\end{equation}
to measure the difference for every model between the computed intensity 
distribution and the best fit to those intensities for each limb-darkening law.
                     Unfortunately, as we showed in Paper~1, the computed 
error depends on how the models are sampled and the number of intensity points.
If one fits intensity profiles for $\mu$-points near the center of the stellar 
disk then the limb-darkening coefficients and predicted errors differ  from 
limb-darkening coefficients and errors predicted from a sample of $\mu$-points 
near the edge of the stellar disk.  However, we can predict the relative 
quality of fits as a function of geometry.  We show in Fig.~\ref{f5} the 
predicted errors for each limb-darkening law as a function of effective 
temperature.

As expected, Fig.~\ref{f5} shows that all six limb-darkening laws fit the 
plane-parallel model atmosphere intensity 
profiles better than intensity profiles from spherical models.
The definition of plane-parallel radiative transfer \citep{Feautrier1964} 
assumes that $I(\mu) \propto e^{-\tau/\mu}$, where $\tau$ is the monochromatic 
optical depth.  As 
$\mu \rightarrow 0$, then $I(\mu) \rightarrow 0$, i.e. the intensity and the 
derivative of the intensity, $\mathrm{d}I/\mathrm{d}\mu$, both change monotonically.  These 
properties allow simple limb-darkening laws to fit plane-parallel model 
intensity profiles well.

For spherically symmetric model atmospheres the radiative transfer is 
calculated for a set of rays along the line-of-sight between the observer and 
points on the stellar disk. The rays nearer the center of the stellar disk come
from depths that are assumed to be infinitely optically thick.  The rays 
farther from the center of the stellar disk penetrate to depths where the 
optical depth is assumed never to reach infinity \citep{Rybicki1971, 
Lester2008}, although the rays can reach extremely large optical depths.  Rays 
located toward the limb of the star can penetrate the tenuous outer atmosphere,
never reaching large optical depths.  As a result, the computed intensity 
profiles have a point of inflection (see Fig.~\ref{f0}) where the intensity 
derivative, $VI/d\mu$, is not changing monotonically, which prevents the 
simple limb-darkening laws from fitting as well.
 
While it is difficult to draw conclusions from the predicted errors, we can 
reliably state that the linear and exponential limb-darkening laws do not fit 
the spherical model atmospheres. The predicted errors for those limb-darkening 
laws range from $0.05$ to $0.2$ and are significantly greater than the errors 
for the fits to plane-parallel models.  The best-fitting relations are the 
square-root law and the four-parameter limb-darkening law of \cite{Claret2000},
which have errors less than $0.08$. 
 
Another thing to note is that based on fits to plane-parallel model 
atmospheres, \cite{Diaz1995} suggested that the square-root law is more 
adequate for fitting hotter stars ($T_{\rm{eff}} >8000~$K), although they were 
unclear which law is preferred for cooler stars.  
For spherical model atmospheres we find that the predicted errors for the 
square-root limb-darkening law are less than the errors for the quadratic law,
making the former the clear preference.
Also, the quadratic limb-darkening law is of particular interest 
because it is the most commonly used limb-darkening law for analyzing 
planetary transit observations \citep{Mandel2003}.  However, numerous comparisons of 
quadratic limb-darkening laws fit directly to observations and those fit to 
model stellar atmospheres suggest disagreement for a number of cases 
\citep{Howarth2011}.  The results presented here suggest it may be advantageous
to consider fitting transit observations with a square-root limb-darkening law 
or the more accurate four-parameter limb-darkening law.
 
\section{Gravity-darkening coefficients}
\cite{Claret2011} computed wavelength-dependent gravity-darkening coefficients 
from {\it Atlas} plane-parallel model stellar atmospheres based on the 
analytic relation developed by \cite{Bloemen2011}.  In Paper~1 we used this 
same prescription for both plane-parallel and spherically symmetric model 
stellar atmospheres to compute 
gravity-darkening coefficients for cool giant stars, and we found that model 
geometry played a negligible role in determining gravity-darkening 
coefficients except for $T_\mathrm{eff} < 4000$~K.  
At the cooler effective temperatures, the 
spherically symmetric model gravity-darkening coefficients are predicted to be 
vary significantly, and are up to an order-of-magnitude greater than those 
predicted from plane-parallel model atmospheres.

We repeat that analysis here for our higher gravity model stellar atmospheres. 
As described by \cite{Bloemen2011}, the gravity-darkening coefficient, 
$y(\lambda)$ for a star is
\begin{equation}\label{eq:grav}
y(\lambda) = \left(\frac{\partial \ln I(\lambda)}{\partial \ln g}\right)_{T_{\rm{eff}}} + \left(\frac{d\ln T_{\rm{eff}}}{d\ln g}\right)\left(\frac{\partial \ln I(\lambda)}{\partial \ln T_{\rm{eff}} }\right)_g.
\end{equation}
As described in Paper~1, \citet{vonZeipel1924} showed that 
$T_\mathrm{eff} \sim (g_\mathrm{eff})^{\beta_1/4}$, 
where $\beta_1 \equiv d\ln T_{\rm{eff}}/d\ln g$.  As previously, we assume 
$\beta_1 = 0.2$ for models with $T_{\rm{eff}} < 7500$~K and $\beta_1 = 1$ 
otherwise.  Using these constant values for $\beta_1$ provides only a 
limited analysis of the gravity-darkening because $\beta_1$ is a function of 
effective temperature, but assuming these two values does enable us to gain 
some perspective on the role of model atmosphere geometry.
The other terms are the partial derivatives of the 
wavelength-dependent intensity with respect to gravity and effective 
temperature, respectively.

\begin{figure*}[t]
\begin{center}
\includegraphics[width=0.5\textwidth]{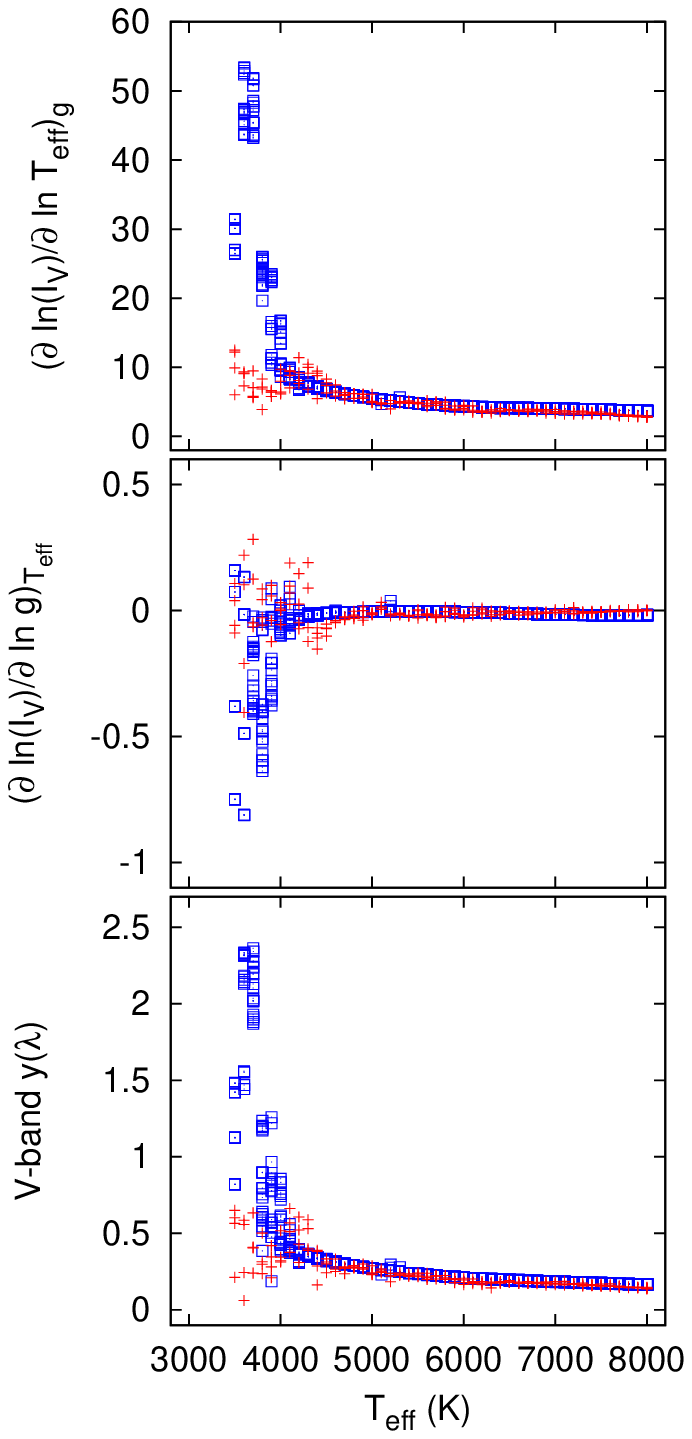}\includegraphics[width=0.5\textwidth]{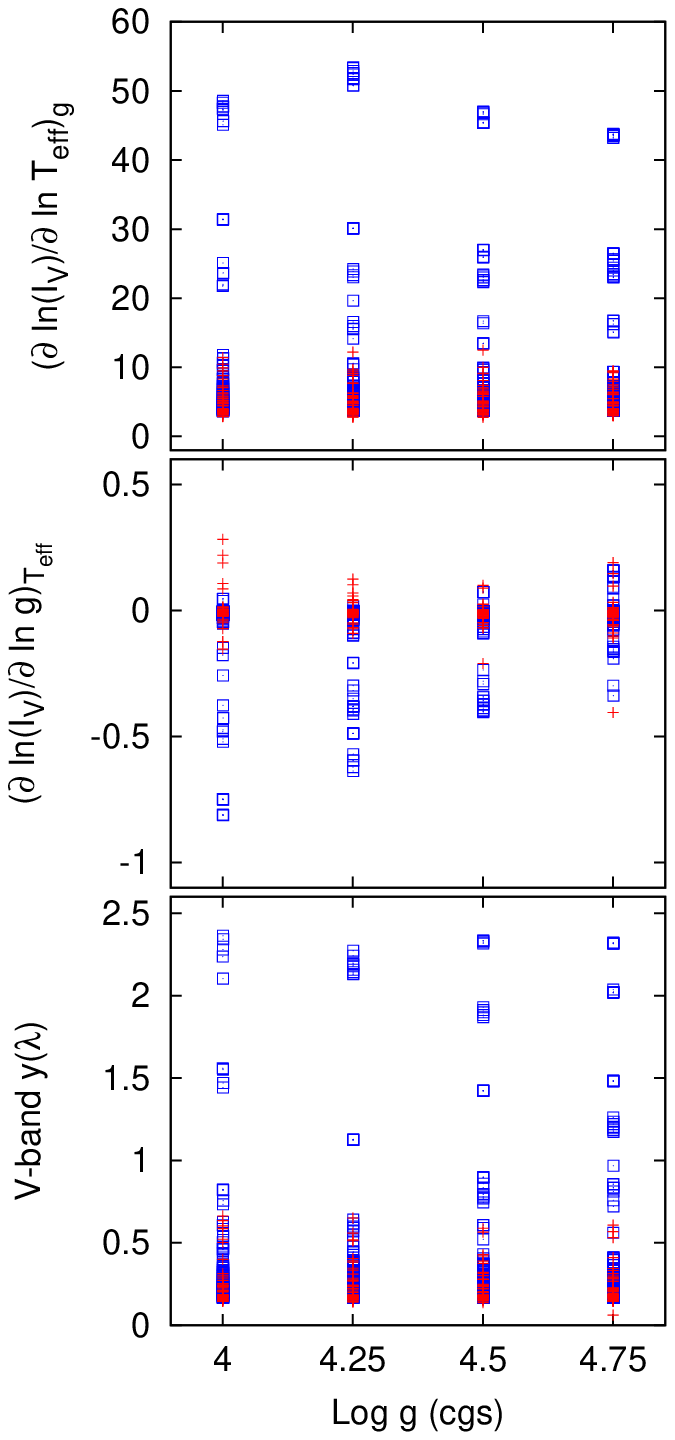}
\end{center}
\caption{$V$-band central intensity derivatives and gravity-darkening 
coefficients as function of effective temperature (left) and gravity 
(right) computed from plane-parallel (red crosses) and 
spherically symmetric (blue squares) model stellar atmospheres.}\label{f6}
\end{figure*}

We compute the two intensity derivatives and predicted gravity-darkening 
coefficients for our grids of plane-parallel and spherically symmetric model 
atmospheres and plot the predicted values in Fig.~\ref{f6} for the {\em Kepler}
waveband. The predicted derivatives and gravity-darkening coefficients are 
similar to those computed in Paper~1, for which there is little difference 
between spherically symmetric and plane-parallel model predictions for 
effective temperatures greater than $4000~$K.  The spherical and planar 
predictions then diverge for cooler effective temperatures. However, the range 
of values for the spherical model predictions is less for the higher gravity 
models explored in this work relative to the lower gravity models studied in 
Paper~1.

\section{Interferometric angular diameter corrections}
Interferometry provides precise measurements of stellar angular diameters.  
However, stellar interferometry measures the combination of angular diameter 
and intensity profile and the two quantities are degenerate.  One route to 
break the degeneracy is to assume a uniform intensity profile and measure the 
uniform-disk angular diameter.  The limb-darkened angular diameter can then be 
predicted from the uniform-disk angular diameter using corrections computed 
from stellar atmosphere models \citep{Davis2000}. 

Another technique for measuring limb-darkened  angular diameters is to assume 
a simple limb-darkening law and coefficients from model stellar atmospheres to 
fit the interferometric observations \citep[e.g][]{Boyajian2012}.  However, this 
technique might also predict incorrect angular diameters because plane-parallel
model atmospheres are typically used for fitting limb-darkening coefficients.  
We can assess 
           the potential error of assuming plane-parallel limb-darkening 
coefficients to fit the angular diameter
by comparing predicted angular diameter corrections from spherically symmetric 
model stellar atmospheres with those from plane-parallel models.
\begin{figure*}[t]
\begin{center}
\includegraphics[width=0.5\textwidth]{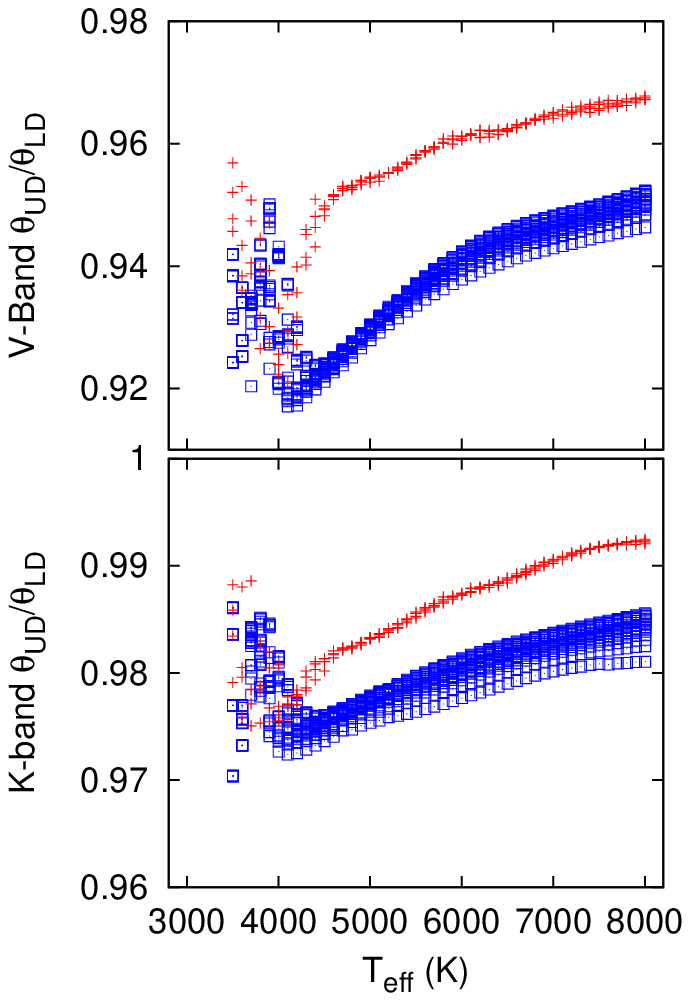}\includegraphics[width=0.5\textwidth]{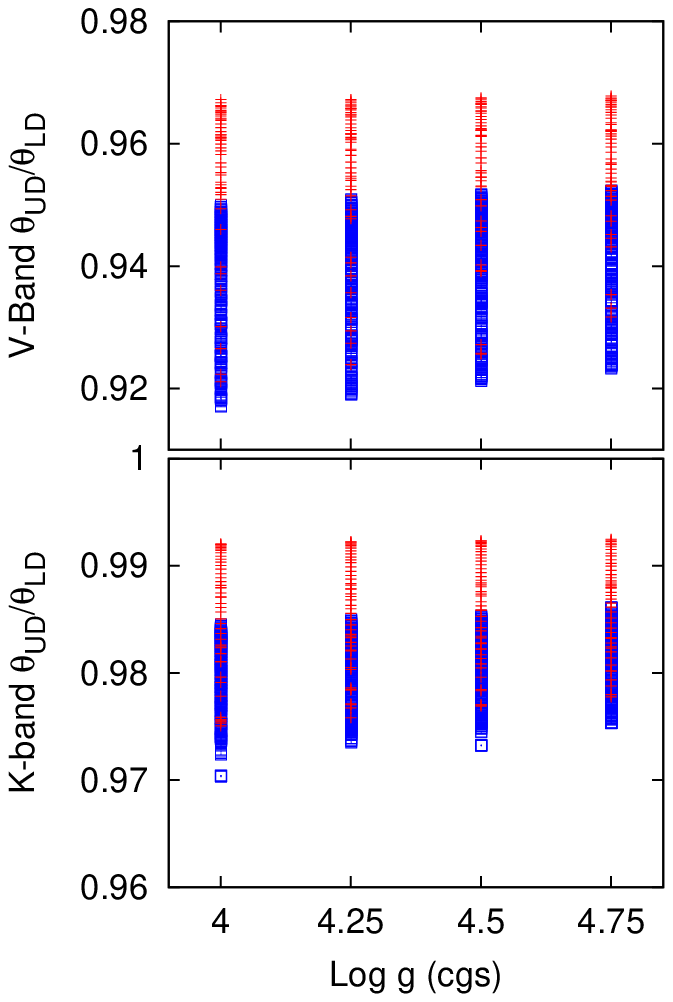}
\end{center}
\caption{Interferometric angular diameter correction computed in $V$-band (top)
and $K$-band (bottom) as functions of effective temperature (left) and gravity 
(right). Corrections computed from plane-parallel model atmospheres are 
denoted with red x's and spherically symmetric models blue squares.}\label{f7}
\end{figure*}

In Fig.~\ref{f7} we plot the $V$- and $K$-band angular diameter corrections as 
a function of effective temperature and gravity for both spherical and planar 
model atmospheres.  The $V$-band corrections vary from $0.93$ to $0.97$ for the 
plane-parallel model atmospheres and from $0.92$ to $0.95$ for spherical models. 
     The difference is more apparent if one considers stellar atmospheres 
with        $T_{\rm{eff}} > 4500$~K, where the difference between 
plane-parallel and spherical model corrections is about $0.01$ to $0.02$.  
This suggests that employing plane-parallel model corrections for measuring 
stellar angular diameters from interferometric observations will lead to a 
$1$ to $2\%$ underestimate of the angular diameter.

Similarly, the $K$-band corrections also vary as a function of model atmosphere
geometry; plane-parallel models suggest values of 
$\theta_{\rm{UD}}/\theta_{\rm{LD}} = 0.98$ to $0.99$ while spherical models 
suggest $\theta_{\rm{UD}}/\theta_{\rm{LD}} = 0.97$ to $0.985$.  Again, using 
plane-parallel model corrections to fit $K$-band interferometric observations 
will underestimate the actual angular diameter by about 1\%.  Thus, for 
precision measurements of angular diameters, hence fundamental stellar 
parameters from optical interferometry, one should employ more 
physically representative spherical model atmospheres.  This appears to be the 
case even for main sequence stars with large gravities and small atmospheric 
extensions.

\section{Summary}
In this work, we followed up on the study of Paper~1 to measure how model 
stellar atmosphere geometry affects predicted limb-darkening coefficients, 
gravity-darkening coefficients and interferometric angular diameter corrections
for main sequence FGK dwarf stars.  As in Paper~1, we find significant 
differences between predictions from plane-parallel and spherically symmetric 
model atmospheres computed with the {\em Atlas/SAtlas} codes.  The results in 
this article are surprising because geometry is believed to be not important for stars
with smaller atmospheric extension, i.e. main sequence stars with $\log g \ge 4$. 
                              As atmospheric extension gets smaller, defined 
as the ratio of the atmospheric depth to  stellar radius, then it is expected 
that a spherical model atmosphere should appear more and more like a 
plane-parallel model atmosphere.  However, even for small atmospheric extension
models, we find differences in predicted intensity profiles, hence differences 
in limb-darkening and angular diameter corrections.

As in Paper~1, there is negligible difference between gravity-darkening 
coefficients predicted from planar and spherical model atmospheres.  This is 
because gravity-darkening coefficients depend heavily on the central intensity of the 
star, not the entire intensity profile.  The central intensity is 
approximately a function of the effective temperature at $\tau_{\rm{Ross}} =1$ 
according to the Eddington-Barbier relation \citep{Mihalas1978}. 
\cite{Lester2008} showed that the atmospheric temperature structure computed 
from plane-parallel and spherically symmetric model atmospheres for the same 
effective temperature and gravity primarily differs closer to the surface, 
$\tau_{\rm{Ross}} < 2/3$, and converges as $\tau \rightarrow \infty$. Because 
the computed temperatures at depth are very     similar for the two geometries,
     the central intensity is also similar for both model geometries, making 
the gravity-darkening coefficients     insensitive to model geometry. However, 
geometry is important for stars with   $T_{\rm{eff}} < 4000~$K, which is due 
to differences in the opacity structure and convection, which lead to changes in the 
temperature structure.

Angular diameter corrections do vary as a function of geometry. The corrections
                account for the degeneracy between the intensity profile and 
limb-darkened angular diameter in modeling interferometric observations.  
Therefore, differences between the intensity profiles of plane-parallel and spherically symmetric model 
stellar atmospheres                   lead directly to differences between 
predicted angular diameter corrections.  We find that spherically symmetric 
model corrections are about $1$ to $2\%$ smaller than planar model corrections
for the main sequence stars analyzed here.

Similarly, we computed limb-darkening coefficients for six different 
limb-darkening laws.   As in Paper~1, we find that the linear law is least 
consistent with predicted intensity profiles and that the four-parameter law 
is best.  We also find that the commonly used quadratic limb-darkening law does 
not fit spherically symmetric model atmosphere intensity profiles as precisely 
as the similar square-root or four-parameter limb-darkening laws.  This suggests
that as planetary-transit observations become increasingly precise, the 
four-parameter law combined with the more physically representative 
spherically symmetric model stellar atmospheres will be more appropriate for 
fitting observations or, better still, using intensity profiles directly.

The angular-diameter corrections, limb-darkening and gravity-darkening 
coefficients are publicly available as online tables. Each table has the 
format $T_\mathrm{eff}$~(K), $\log g$ and $M~(M_\odot)$ and then the appropriate variables for each waveband, such as 
linear limb-darkening coefficients. Tables for plane-parallel model fits do not include mass. Tables of gravity-darkening coefficients 
also contain values of the intensity derivatives with respect to gravity and 
effective temperature. For plane-parallel models, values of mass, radius and 
luminosity are presented as zero in the tables. We list the properties of 
these tables in Table~\ref{t1} that are available from CDS. Tabulated grids of 
the model atmosphere intensity profiles used in this work are also available.

\begin{table}[t]
\caption{Summary of limb-darkening coefficient, gravity-darkening 
coefficient and interferometric angular diameter correction tables 
found online.}\label{t1}
\begin{center}
\begin{tabular}{lll}
\hline
\hline
Name & Geometry & Type \\
\hline
Table2  & Spherical & Linear Limb Darkening Eq.~\ref{eq:linear} \\
Table3  & Spherical & Quadratic Limb Darkening Eq.~\ref{eq:quad} \\
Table4  & Spherical & Square Root Limb Darkening Eq.~\ref{eq:root} \\
Table5  & Spherical & Four-parameter Limb Darkening Eq.~\ref{eq:4-p} \\
Table6  & Spherical & Exponential Limb Darkening Eq.~\ref{eq:exp} \\
Table7  & Spherical & Logarithmic Limb Darkening Eq.~\ref{eq:ln} \\
Table8  & Planar    & Linear Limb Darkening Eq.~\ref{eq:linear} \\
Table9  & Planar    & Quadratic Limb Darkening Eq.~\ref{eq:quad} \\
Table10 & Planar    & Square Root Limb Darkening Eq.~\ref{eq:root} \\
Table11 & Planar    & Four-parameter Limb Darkening Eq.~\ref{eq:4-p} \\
Table12 & Planar    & Exponential Limb Darkening Eq.~\ref{eq:exp} \\
Table13 & Planar    & Logarithmic Limb Darkening Eq.~\ref{eq:ln} \\
Table14 & Spherical & Gravity Darkening \\
Table15 & Planar    & Gravity Darkening \\
Table16 & Spherical & Angular Diameter Corrections \\
Table17 & Planar & Angular Diameter Corrections \\
\hline
\end{tabular}
\end{center}
\note{Tables listed here can be retrieved electronically from the CDS.}
\end{table}

\acknowledgements

This work has been supported by a research grant from the Natural 
Sciences and Engineering Research Council of Canada, the Alexander von Humboldt Foundation and NSF grant (AST-0807664).

\bibliographystyle{aa} 

\bibliography{ld4}

\begin{thebibliography}{35}
\expandafter\ifx\csname natexlab\endcsname\relax\def\natexlab#1{#1}\fi

\bibitem[{{Al-Naimiy}(1978)}]{Al-naimiy1979}
{Al-Naimiy}, H.~M. 1978, \apss, 53, 181

\bibitem[{{An} {et~al.}(2002){An}, {Albrow}, {Beaulieu}, {Caldwell}, {DePoy},
  {Dominik}, {Gaudi}, {Gould}, {Greenhill}, {Hill}, {Kane}, {Martin},
  {Menzies}, {Pogge}, {Pollard}, {Sackett}, {Sahu}, {Vermaak}, {Watson}, \&
  {Williams}}]{An2002}
{An}, J.~H., {Albrow}, M.~D., {Beaulieu}, J.-P., {et~al.} 2002, \apj, 572, 521

\bibitem[{{Auvergne} {et~al.}(2009){Auvergne}, {Bodin}, {Boisnard}, {Buey},
  {Chaintreuil}, {Epstein}, {Jouret}, {Lam-Trong}, {Levacher}, {Magnan},
  {Perez}, {Plasson}, {Plesseria}, {Peter}, {Steller}, {Tiph{\`e}ne}, {Baglin},
  {Agogu{\'e}}, {Appourchaux}, {Barbet}, {Beaufort}, {Bellenger}, {Berlin},
  {Bernardi}, {Blouin}, {Boumier}, {Bonneau}, {Briet}, {Butler}, {Cautain},
  {Chiavassa}, {Costes}, {Cuvilho}, {Cunha-Parro}, {de Oliveira Fialho},
  {Decaudin}, {Defise}, {Djalal}, {Docclo}, {Drummond}, {Dupuis}, {Exil},
  {Faur{\'e}}, {Gaboriaud}, {Gamet}, {Gavalda}, {Grolleau}, {Gueguen},
  {Guivarc'h}, {Guterman}, {Hasiba}, {Huntzinger}, {Hustaix}, {Imbert},
  {Jeanville}, {Johlander}, {Jorda}, {Journoud}, {Karioty}, {Kerjean},
  {Lafond}, {Lapeyrere}, {Landiech}, {Larqu{\'e}}, {Laudet}, {Le Merrer},
  {Leporati}, {Leruyet}, {Levieuge}, {Llebaria}, {Martin}, {Mazy}, {Mesnager},
  {Michel}, {Moalic}, {Monjoin}, {Naudet}, {Neukirchner}, {Nguyen-Kim},
  {Ollivier}, {Orcesi}, {Ottacher}, {Oulali}, {Parisot}, {Perruchot},
  {Piacentino}, {Pinheiro da Silva}, {Platzer}, {Pontet}, {Pradines},
  {Quentin}, {Rohbeck}, {Rolland}, {Rollenhagen}, {Romagnan}, {Russ}, {Samadi},
  {Schmidt}, {Schwartz}, {Sebbag}, {Smit}, {Sunter}, {Tello}, {Toulouse},
  {Ulmer}, {Vandermarcq}, {Vergnault}, {Wallner}, {Waultier}, \&
  {Zanatta}}]{Auvergne2009}
{Auvergne}, M., {Bodin}, P., {Boisnard}, L., {et~al.} 2009, \aap, 506, 411

\bibitem[{{Barros} {et~al.}(2012){Barros}, {Pollacco}, {Gibson}, {Keenan},
  {Skillen}, \& {Steele}}]{Barros2012}
{Barros}, S.~C.~C., {Pollacco}, D.~L., {Gibson}, N.~P., {et~al.} 2012, \mnras,
  419, 1248

\bibitem[{{Bass} {et~al.}(2012){Bass}, {Orosz}, {Welsh}, {Windmiller}, {Ames
  Gregg}, {Fetherolf}, {Wade}, \& {Quinn}}]{Bass2012}
{Bass}, G., {Orosz}, J.~A., {Welsh}, W.~F., {et~al.} 2012, \apj, 761, 157

\bibitem[{{Bessell}(2005)}]{Bessell2005}
{Bessell}, M.~S. 2005, \araa, 43, 293

\bibitem[{{Bloemen} {et~al.}(2011){Bloemen}, {Marsh}, {{\O}stensen},
  {Charpinet}, {Fontaine}, {Degroote}, {Heber}, {Kawaler}, {Aerts}, {Green},
  {Telting}, {Brassard}, {G{\"a}nsicke}, {Handler}, {Kurtz}, {Silvotti}, {van
  Grootel}, {Lindberg}, {Pursimo}, {Wilson}, {Gilliland}, {Kjeldsen},
  {Christensen-Dalsgaard}, {Borucki}, {Koch}, {Jenkins}, \&
  {Klaus}}]{Bloemen2011}
{Bloemen}, S., {Marsh}, T.~R., {{\O}stensen}, R.~H., {et~al.} 2011, \mnras,
  410, 1787

\bibitem[{{Boyajian} {et~al.}(2012){Boyajian}, {von Braun}, {van Belle},
  {McAlister}, {ten Brummelaar}, {Kane}, {Muirhead}, {Jones}, {White},
  {Schaefer}, {Ciardi}, {Henry}, {L{\'o}pez-Morales}, {Ridgway}, {Gies}, {Jao},
  {Rojas-Ayala}, {Parks}, {Sturmann}, {Sturmann}, {Turner}, {Farrington},
  {Goldfinger}, \& {Berger}}]{Boyajian2012}
{Boyajian}, T.~S., {von Braun}, K., {van Belle}, G., {et~al.} 2012, \apj, 757,
  112

\bibitem[{{Claret}(2000)}]{Claret2000}
{Claret}, A. 2000, \aap, 363, 1081

\bibitem[{{Claret} \& {Bloemen}(2011)}]{Claret2011}
{Claret}, A. \& {Bloemen}, S. 2011, \aap, 529, A75

\bibitem[{{Claret} \& {Hauschildt}(2003)}]{Claret2003}
{Claret}, A. \& {Hauschildt}, P.~H. 2003, \aap, 412, 241

\bibitem[{{Claret} {et~al.}(2012){Claret}, {Hauschildt}, \&
  {Witte}}]{Claret2012}
{Claret}, A., {Hauschildt}, P.~H., \& {Witte}, S. 2012, \aap, 546, A14

\bibitem[{{Claret} {et~al.}(2013){Claret}, {Hauschildt}, \&
  {Witte}}]{Claret2013}
{Claret}, A., {Hauschildt}, P.~H., \& {Witte}, S. 2013, \aap, 552, A16

\bibitem[{{Croll} {et~al.}(2011){Croll}, {Albert}, {Jayawardhana},
  {Miller-Ricci Kempton}, {Fortney}, {Murray}, \& {Neilson}}]{Croll2011}
{Croll}, B., {Albert}, L., {Jayawardhana}, R., {et~al.} 2011, \apj, 736, 78

\bibitem[{{Davis} {et~al.}(2000){Davis}, {Tango}, \& {Booth}}]{Davis2000}
{Davis}, J., {Tango}, W.~J., \& {Booth}, A.~J. 2000, \mnras, 318, 387

\bibitem[{{Diaz-Cordoves} {et~al.}(1995){Diaz-Cordoves}, {Claret}, \&
  {Gimenez}}]{Diaz1995}
{Diaz-Cordoves}, J., {Claret}, A., \& {Gimenez}, A. 1995, \aaps, 110, 329

\bibitem[{{Feautrier}(1964)}]{Feautrier1964}
{Feautrier}, P. 1964, Comptes Rendus Academie des Sciences (serie non
  specifiee), 258, 3189

\bibitem[{{Fields} {et~al.}(2003){Fields}, {Albrow}, {An}, {Beaulieu},
  {Caldwell}, {DePoy}, {Dominik}, {Gaudi}, {Gould}, {Greenhill}, {Hill},
  {J{\o}rgensen}, {Kane}, {Martin}, {Menzies}, {Pogge}, {Pollard}, {Sackett},
  {Sahu}, {Vermaak}, {Watson}, {Williams}, {Glicenstein}, {Hauschildt}, \&
  {PLANET Collaboration}}]{Fields2003}
{Fields}, D.~L., {Albrow}, M.~D., {An}, J., {et~al.} 2003, \apj, 596, 1305

\bibitem[{{Hauschildt} {et~al.}(1999){Hauschildt}, {Allard}, \&
  {Baron}}]{Hauschildt1998}
{Hauschildt}, P.~H., {Allard}, F., \& {Baron}, E. 1999, \apj, 512, 377

\bibitem[{{Howarth}(2011{\natexlab{a}})}]{Howarth2011a}
{Howarth}, I.~D. 2011{\natexlab{a}}, \mnras, 413, 1515

\bibitem[{{Howarth}(2011{\natexlab{b}})}]{Howarth2011}
{Howarth}, I.~D. 2011{\natexlab{b}}, \mnras, 418, 1165

\bibitem[{{Johnson} \& {Morgan}(1953)}]{Johnson1953}
{Johnson}, H.~L. \& {Morgan}, W.~W. 1953, \apj, 117, 313

\bibitem[{{Koch} {et~al.}(2004){Koch}, {Borucki}, {Dunham}, {Geary},
  {Gilliland}, {Jenkins}, {Latham}, {Bachtell}, {Berry}, {Deininger}, {Duren},
  {Gautier}, {Gillis}, {Mayer}, {Miller}, {Shafer}, {Sobeck}, {Stewart}, \&
  {Weiss}}]{Koch2004}
{Koch}, D.~G., {Borucki}, W., {Dunham}, E., {et~al.} 2004, in Society of
  Photo-Optical Instrumentation Engineers (SPIE) Conference Series, Vol. 5487,
  Society of Photo-Optical Instrumentation Engineers (SPIE) Conference Series,
  ed. J.~C. {Mather}, 1491--1500

\bibitem[{{Kurucz}(1979)}]{Kurucz1979}
{Kurucz}, R.~L. 1979, \apjs, 40, 1

\bibitem[{{Lester} \& {Neilson}(2008)}]{Lester2008}
{Lester}, J.~B. \& {Neilson}, H.~R. 2008, \aap, 491, 633

\bibitem[{{Mandel} \& {Agol}(2002)}]{Mandel2003}
{Mandel}, K. \& {Agol}, E. 2002, \apjl, 580, L171

\bibitem[{{Mihalas}(1978)}]{Mihalas1978}
{Mihalas}, D. 1978, {Stellar atmospheres /2nd edition/} (San Francisco:
  W.~H.~Freeman and Co.)

\bibitem[{{Neilson} \& {Lester}(2011)}]{Neilson2011}
{Neilson}, H.~R. \& {Lester}, J.~B. 2011, \aap, 530, A65

\bibitem[{{Neilson} \& {Lester}(2012)}]{Neilson2012a}
{Neilson}, H.~R. \& {Lester}, J.~B. 2012, \aap, 544, A117

\bibitem[{{Neilson} \& {Lester}(2013)}]{Neilson2012}
{Neilson}, H.~R. \& {Lester}, J.~B. 2013, \aap, 554, A98

\bibitem[{{Rybicki}(1971)}]{Rybicki1971}
{Rybicki}, G.~B. 1971, \jqsrt, 11, 589

\bibitem[{{Sing}(2010)}]{Sing2010}
{Sing}, D.~K. 2010, \aap, 510, A21

\bibitem[{{van Hamme}(1993)}]{vanhamme1993}
{van Hamme}, W. 1993, \aj, 106, 2096

\bibitem[{{von Zeipel}(1924)}]{vonZeipel1924}
{von Zeipel}, H. 1924, \mnras, 84, 665

\bibitem[{{Wade} \& {Rucinski}(1985)}]{Wade1985}
{Wade}, R.~A. \& {Rucinski}, S.~M. 1985, \aaps, 60, 471

\end{thebibliography}

\end{document}